\documentclass[a4paper,twoside]{article}

\usepackage{epsfig}
\usepackage{subcaption}
\usepackage{calc}
\usepackage{amssymb}
\usepackage{amstext}
\usepackage{amsmath}
\usepackage{amsthm}
\usepackage{multicol}
\usepackage{pslatex}
\usepackage{apalike}
\usepackage{physics}
\usepackage{algorithm2e}
\usepackage[bottom]{footmisc}
\usepackage{SCITEPRESS}     
\usepackage{tikz}
\usetikzlibrary{positioning}

\begin{document}

\title{Benchmarking Quantum Reinforcement Learning}

\author{\authorname{Georg Kruse\sup{1,2}, Rodrigo Coelho\sup{1}, Andreas Rosskopf\sup{1}, Robert Wille\sup{2}, and Jeanette Miriam Lorenz\sup{3,4}}
\affiliation{\sup{1}Fraunhofer IISB, Erlangen, Germany}
\affiliation{\sup{2}Technical University Munich, Germany}
\affiliation{\sup{3}Ludwig Maximilian University, Germany}
\affiliation{\sup{4}Fraunhofer IKS, Munich, Germany}
\email{\{georg.kruse, rodrigo.coelho, andreas.rosskopf\}@iisb.fraunhofer.de}
}


\abstract{Quantum Reinforcement Learning (QRL) has emerged as a promising research field, leveraging the principles of quantum mechanics to enhance the performance of reinforcement learning (RL) algorithms. However, despite its growing interest, QRL still faces significant challenges. It is still uncertain if QRL can show any advantage over classical RL beyond artificial problem formulations. Additionally, it is not yet clear which  streams of QRL research show the greatest potential. The lack of a unified benchmark and the need to evaluate the reliance on quantum principles of QRL approaches are pressing questions. This work aims to address these challenges by providing a comprehensive comparison of three major QRL classes: Parameterized Quantum Circuit based QRL (PQC-QRL) (with one policy gradient (QPG) and one Q-Learning (QDQN) algorithm), Free Energy based QRL (FE-QRL), and Amplitude Amplification based QRL (AA-QRL). We introduce a set of metrics to evaluate the QRL algorithms on the widely applicable benchmark of gridworld games. Our results provide a detailed analysis of the strengths and weaknesses of the QRL classes, shedding light on the role of quantum principles in QRL and paving the way for future research in this field.}

\onecolumn \maketitle \normalsize \setcounter{footnote}{0} \vfill

\section{Introduction}

Quantum Reinforcement Learning (QRL) has gained significant attention in recent years. Various approaches have been proposed to leverage the principles of quantum mechanics to enhance the performance of classical reinforcement learning (RL) algorithms. Initially, QRL research focused on amplitude amplification techniques applied to tasks like gridworld navigation \cite{dong2008quantum}. The emergence of quantum annealers like D-Wave led to the development of free energy based learning using Quantum Boltzmann Machines (QBMs) \cite{Crawford.2018}. Most recently, the widespread use of parameterized quantum circuits (PQC), often referred to as quantum neural networks \cite{abbas2021power}, has led to a wider range of applications of QRL algorithms.

Despite its growing interest, QRL still faces significant challenges. While it is still uncertain whether QRL can outperform classical RL, it is also unclear which QRL approach holds the most promise. Until now, only a limited number of works (e.g. \cite{Neumann.2023}) have compared the various classes of QRL algorithms against each other and no unified benchmarks have been proposed. A critical gap in QRL research is the lack of studies examining whether any performance enhancements are due to quantum properties: \cite{bowles2024better} have raised questions about the reliance of quantum models on entanglement and superposition. This work seeks to contribute to this discourse by examining whether QRL algorithms genuinely rely on their quantum parts, or if algorithms without them can achieve similar results.

Hence, we provide a comprehensive comparison of three of the most widely spread QRL classes: Parameterized Quantum Circuit based QRL (PQC-QRL) (with one policy gradient (QPG) and one Q-Learning (QDQN) algorithm), Free Energy based QRL (FE-QRL), and Amplitude Amplification based QRL (AA-QRL), which we will briefly introduce in Section \ref{algorithms}. We compare these QRL approaches using a series of metrics, including the number of required quantum circuit executions and the estimated quantum clock time. In addition to these  metrics we will also investigate whether or not the performance relies on the quantum properties of the quantum approaches.

By establishing a benchmark which is applicable to various QRL algorithms, this paper aims to serve as a resource for future QRL research to facilitate a clearer understanding of the relative merits and challenges of each quantum approach. Through a detailed analysis of performance metrics and quantum properties, we aim to guide the development of novel QRL algorithms.

\section{Preliminaries} \label{algorithms}

At their core, all RL algorithms, whether classical or quantum, share a common structure centered around the interaction between an agent and its environment. The agent, responsible for making decisions, consists of a function approximator that learns through interactions with its environment - the external surroundings that influence and respond to the agent's actions. The agent's ultimate goal is to develop a strategy that maximizes the reward it receives from the environment.

Most RL environments are modeled as Markov Decision Processes (MDPs). An MDP is characterized by its state space $S$, its action space $A$, a state transition probability function $P$, denoting the probability of transitioning at time step $t$ from state $s_t$ to the next state $s_{t+1}$ after taking action $a_t$, and a reward function $R$, which quantifies the immediate value of each state-action combination. This reward mechanism serves as the learning signal, guiding the agent towards optimal behavior through the maximization of cumulative rewards.

In the field of QRL, various classes of algorithms have been proposed. While covering all classes is beyond the scope of this work, we will focus on the most established ones that can be applied to the most generic benchmark case, namely gridworld games (we will motivate the choice of gridworld games in Section \ref{sec:benchmarks}). We refer the reader to a comprehensive overview of the field of QRL to the review by \cite{meyer2022survey}. In this work, we will focus on three classes of QRL, which we will briefly introduce in the following subsections.

\subsection{Parameterized Quantum Circuit based QRL}

In Deep Reinforcement Learning (DRL), deep neural networks (DNNs) serve as powerful function approximators. In the stream of research which we will refer to as PQC-QRL, the DNNs are replaced with PQCs. This approach has gained significant attention among researchers due to its simplicity and natural similarity to classical RL methods, leading to numerous implementations with varying circuit designs \cite{coelho2024vqc} \cite{kruse2024hamiltonian}.
However, the influence of the chosen ansatz remains poorly understood, emphasizing the critical need for systematic benchmarking efforts. Current research often builds upon the hardware-efficient ansatz (HEA), which has been initially used by \cite{chen2020variational} and  \cite{jerbi2021parametrized} and later improved upon, notably by  \cite{skolik2022quantum} with data re-uploading and trainable output scaling parameters.

\begin{figure}[h] 
    \centering
    \includegraphics[width=\linewidth]{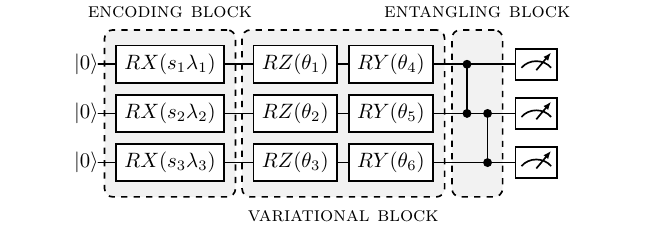}
    \caption{
    A single-layer PQC  $U_{\theta, \lambda}(s)$ for PQC-QRL is typically composed of three blocks that are repeated in each layer: an \emph{encoding block}, where the features of the state (potentially scaled by trainable parameters $\lambda$) are encoded; a \emph{variational block}, with parameterized quantum gates; and an \emph{entangling block}. However, this structure is flexible, allowing the blocks to be rearranged, combined, or modified as needed. In this work we utilize the depicted ansatz as proposed by \cite{skolik2022quantum}.} \label{fig:vqc_hea}
\end{figure}

The evaluation of various ansatz design choices would be beyond the scope of this work and has also been conducted in previous works by \cite{druagan2022quantum} and \cite{kruse2023variational}. Instead, our evaluation aims to establish a benchmark across various algorithmic approaches that future studies can build upon and progressively enhance. Hence, in this work, we use the ansatz proposed by \cite{skolik2022quantum}. Similarly to classical RL, quantum implementations commonly utilize Policy Gradient (PG), Q-Learning (in the form of DQNs), and actor-critic approaches like Proximal Policy Optimization (PPO) as training algorithms. Our investigation focuses on analyzing the performance of quantum implementations of PG and DQN which we will refer to as QPG and QDQN, respectively. 

\subsubsection{Q-Learning}

Q-Learning is a value-based algorithm that estimates the expected return or value of taking a particular action in a given state. The Q-function is defined by mapping a tuple $(\pi, s, a)$ of a given  policy $\pi$, a current state $s$, and an action $a$ to the expected value of the current and future discounted rewards. This is formally expressed through the action-value function 

\begin{equation}
    Q_\pi (s_t, a_t) = \mathbb{E}_\pi \left[ G_t \mid s_t = s, a_t = a \right] ,
\end{equation}

where $G_t$ denotes the cumulative return at time step $t$. To select the next action in state $s_t$, the action corresponding to the maximal Q-value is selected by $a_t = argmax_a \:Q(s_t, a)$. In order to balance between exploration and exploitation, an $\epsilon$-greedy policy is used, which chooses a random action with probability $1 - \epsilon$ and the action with the highest value otherwise. Typically, $\epsilon$ decays over time to favor exploitation as the algorithm converges. 

In PQC-QRL, the DNN is replaced by a PQC denoted by the unitary $U_{\theta,\lambda}(s)$ as function approximator. A single layer of its ansatz is depicted in Fig. \ref{fig:vqc_hea}. With this PQC, the Q-value of a state-action pair can be estimated by a quantum computer by 

\begin{equation}
    Q(s,a) = \bra{0^{\otimes n}}U_{\theta,\lambda}(s)^{\dagger}O_aU_{\theta,\lambda}(s)\ket{0^{\otimes n}}\cdot w_a
\end{equation}

with trainable circuit parameters $\theta$ and $\lambda$, trainable output scaling $w_a$ and action space dependent observable $O_a$ (which we chose to be Pauli-Z operators for the respective action).
To improve training stability an additional function approximator $\hat{U}_{\theta,\lambda}(s)$ with temporarily fixed weights can be implemented as a target network. These fixed weights are updated at regular intervals of $C$ time steps.

\subsubsection{Policy Gradient}

The PG algorithm is a policy-based algorithm, which directly learns the optimal policy without explicitly estimating a value function. The policy $\pi$ of the agent, which in the quantum case is represented by a unitary $U_{\theta, \lambda}$, is updated such that it maximize the expected cumulative reward $G_t$. At every time step $t$, the agent selects an action $a_t$ in the current state $s_t$ according to a probability distribution defined by the policy $\pi$. Specifically, the probability of choosing action $a$ in state $s$ is given by

\begin{equation}
    \pi_{\theta, \lambda, w}(a|s) = \frac{\bra{0^{\otimes n}}U_{\theta,\lambda}(s)^{\dagger}O_aU_{\theta,\lambda}(s)\ket{0^{\otimes n}}\cdot w_a}
    {\sum_{a'}\bra{0^{\otimes n}}U_{\theta,\lambda}(s)^{\dagger}O_{a'}U_{\theta,\lambda}(s)\ket{0^{\otimes n}}\cdot w_{a'}}.
\end{equation}

For a more detailed description of the PQC-QRL algorithms, we refer the reader to \cite{skolik2023robustness}.

\subsection{Free Energy based QRL}\label{sec:FE-QRL}

A classical Boltzmann Machine (BM) can be viewed as a stochastic neural network with two sets of nodes: visible $v$ and hidden $h$  \cite{ackley1985learning}. Each node represents a binary random variable, and the interactions between these nodes are defined by real-valued weighted edges of an undirected graph. Notably, a Generalized Boltzmann Machine (GBM) allows for connections between any two nodes, offering a highly interconnected structure, while Restricted Boltzmann Machines (RBMs) only allow for connections between the visible nodes $v$ and hidden nodes $h$. Deep Boltzmann Machines (DBM) extend this concept by introducing multiple layers of hidden nodes, allowing for connections only between successive layers.

A clamped DBM is a specialized case of the GBM where all visible nodes $v$ are assigned fixed values $v \in \{0,1\}$. Its classical Hamiltonian, denoted by the index $v$ when the binary values are fixed, is given by 

\begin{equation}\label{eq:bm}
    H^{DBM}_v = -  \sum_{v \in V, h \in H} \theta^{vh} v h -  \sum_{\{hh'\} \subseteq H} \theta^{hh'} hh'
\end{equation}

with trainable weights $\theta^{hh'}$ between the hidden nodes and $\theta^{vh}$ between visible and hidden nodes.
When the binary random variables of $H^{DBM}_v$ are replaced by qubits for each node in the underlying graph and a transverse field $\Gamma$ is added, one arrives at the concept of a clamped Quantum Boltzmann Machine (QBM) \cite{Amin.2018} \cite{kappen2020learning}. This transformation leads to the clamped Hamiltonian formulation

\begin{equation}\label{eq:qbm}
\begin{split}
    H^{QBM}_v = & - \sum_{v \in V, h \in H} \theta^{vh} v \sigma_h^z  \\ & -  \sum_{\{hh'\} \subseteq H} \theta^{hh'} \sigma_h^z\sigma_{h'}^z - \Gamma \sum_{h \in H} \sigma^x_h.
\end{split}
\end{equation}

Here $\sigma^x$ and $\sigma^z$ represent Pauli-X and Pauli-Z operators respectively (and $v \in \{-1, +1\}$). This formulation is called a Transverse Field Ising Model (TFIM) (and $\Gamma$ denotes for the strength of this field) where the transverse field terms are applied only to the hidden units (hence this formulation is sometimes also referred to as a semi transverse QBM \cite{Jerbi.2021}). When the transverse field of the clamped QBM is set to zero, it is equivalent to the clamped classical DBM (Eq. \ref{eq:bm}). 

When using a QBM for FE-QRL, one can use the equilibrium free energy $F(v)$ of a QBM to approximate the Q-function \cite{sallans2004reinforcement} \cite{Jerbi.2021}. For a given fixed assignment of the visible nodes $v$ for a clamped QBM we can calculate $F(v)$ via

\begin{equation}\label{eq:free-energy}
    F(v) = - \frac{1}{\beta} ln\: Z_v = \langle H_v \rangle + \frac{1}{\beta}tr(\rho_v ln \:\rho_v) ,
\end{equation}

with a fixed thermodynamic $\beta = \frac{1}{k_B T}$ (with Boltzmann constant $k_B$ and temperature $T$) and the partition function $Z_v = tr(e^{-\beta H_v})$ and density matrix $\rho_v = \frac{1}{Z_v}e^{-\beta H_v}$ \cite{Crawford.2018}. $\langle H_v \rangle$ represents the expected value of any observable with respect to the Gibbs measure (i.e., the Boltzmann distribution)\cite{levit2017free}

\begin{equation}
    \langle H_v \rangle = \frac{1}{Z_v} tr(H_v e^{-\beta H_v}).
\end{equation}

The negative free energy of a QBM can then be used to approximate the Q-function through the relationship in Eq. \ref{eq:f} for a fixed assignment of state and action $s$ and $a$, which are encoded via the visible nodes $v = \{s,a\}$.

\begin{equation}\label{eq:f}
    Q(s,a) = - F(s,a)
\end{equation}

Using the temporal difference (TD) one-step update rule, the parameters of the QBM can be updated to learn from interactions with the environment. As shown in \cite{levit2017free} and \cite{Crawford.2018}, we obtain:

\begin{equation}
\begin{split}
    \triangle \theta^{vh} =\ & \alpha \ (R_t(s_t, a_t) - \gamma \  F(s_{t+1}, a_{t+1}) \\ 
    &+ F(s_t, a_t)\big)\cdot v \langle \sigma_h^z \rangle ,
\end{split}
\end{equation}

\begin{equation}
\begin{split}
     \triangle \theta^{hh'} =\ & \alpha \ (R_t(s_t, a_t) - \gamma \ F(s_{t+1}, a_{t+1}) \\ &
     + F(s_t, a_t)\big)  
    \cdot  \langle \sigma_h^z  \sigma_{h'}^z  \rangle .
\end{split}
\end{equation}

Here $\alpha$ is the learning rate, $\gamma$ a discount factor and $R_t$ the reward function. We can approximate the expectation values of the observables $\langle \sigma_h^z \rangle $ and $\langle \sigma_h^z \sigma_{h'}^z \rangle $ via sampling from a quantum computer. However, the difficulty of estimating the free energy $F(s,a)$ with a quantum computer remains, as we will discuss in more detail in Section \ref{eval-q}. 

\subsection{Amplitude Amplification based QRL} \label{sec:aa}

The third class of QRL, which we refer to as AA-QRL, was originally proposed by \cite{dong2008quantum}. This method initializes a quantum circuit that encompasses all possible states $s$ and operates by modulating the probability amplitudes for these states according to received rewards through controlled Grover iterations. While the original authors \cite{dong2008quantum} suggested their algorithm could operate in superposition across all possible states, their implementation focused solely on individual state updates. Therefore, this approach is (currently) also considered as quantum inspired RL (QiRL). 

AA-QRL represents an alternative to conventional TD algorithms.
The training of the algorithm begins with $n$ quantum registers (corresponding to $n$ states), where each register is initialized in an equal superposition of $m$ qubits. The $2^m$ possible eigenstates correspond to the available actions.

Following the $TD(0)$ framework, updates of the value function $V$ are performed and the algorithm adjusts the action probabilities in the respective states by applying the Grover operator $L$ times, where $L$ is calculated as $L=int(k \cdot (R_t(s_t,a_t) + V(s_{t+1})))$. The hyperparameter $k$ influences how many Grover iterations are performed, making $L$ proportional to $R_t(s_t,a_t) + V(s_{t+1})$ \cite{dong2008quantum}. This quantum approach differs from classical exploration strategies such as epsilon-greedy or Boltzmann exploration (softmax). Works by \cite{dong2010robust} and \cite{hu2021quantum} have demonstrated that AA-QRL exhibits superior robustness to learning rate variations and initial state conditions compared to traditional RL methods.

\section{How to benchmark QRL}

Benchmarking QRL algorithms requires careful consideration of three  factors: First, a suitable benchmark environment is required, one that can be applied across a wide range of QRL algorithms. On these environments all agents will be given the same amount of maximal environment interactions. We will motivate the choice of benchmark environments for QRL algorithms in Section \ref{sec:benchmarks}. 

Second, performance metrics which align with those used in classical RL need to be introduced (ref. Section  \ref{metrics}). These should incorporate all relevant phenomena of QRL and facilitate future comparisons between classical and quantum agents. However, in this work we will only focus on the comparison between quantum agents.

Third, as demonstrated by \cite{bowles2024better}, it is crucial to investigate whether any observed advantages in QRL stem from quantum principles or other factors. Therefore one needs to establish additional evaluation procedures beyond the metrics of Section \ref{metrics} to investigate these phenomena (ref. Section \ref{eval-q}).

All QRL algorithms introduced in Section \ref{algorithms} have been compared to classical RL agents in various previous studies, some of which are referenced in the corresponding Sections. We therefore do not include this classical comparison in this work. Instead, our goal is to establish a consistent benchmark for different streams of QRL algorithms on which future work can build upon. 

\subsection{Benchmark Environments} \label{sec:benchmarks}

The choice of appropriate benchmark environments for QRL is crucial for meaningful evaluation and comparison of different approaches. In classical RL, the OpenAI Gym (now known as \textit{gymnasium} \cite{towers2024gymnasium}) is a well-established benchmark environment library, offering environments from simple control tasks to complex Atari games. However, only a subset of the \textit{gymnasium} environments are suited for a comparison of the quantum algorithms introduced in Section \ref{algorithms}. In fact, the only type of environment applicable to all introduced QRL algorithms in the \textit{gymnasium} library are the ones with discrete state and action spaces.

\begin{figure}[h]
\centering
\begin{subfigure}{0.22\textwidth}
    \includegraphics[width=0.95\linewidth]{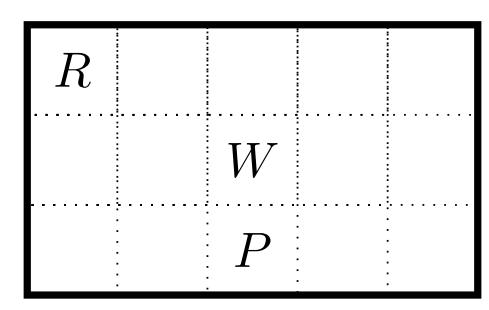}
    \label{fig:first}
\end{subfigure}
\begin{subfigure}{0.22\textwidth}
    \includegraphics[width=0.75\linewidth]{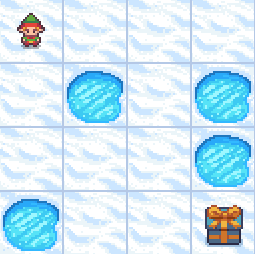}
    \label{fig:second}
\end{subfigure}

\caption{Examples of two commonly used gridworld games: Classical gridworlds with reward $R$, walls $W$ and penalties $P$ as proposed by \cite{sutton1990integrated} and \cite{Crawford.2018} (left). Example of a $4\times4$ instance of the \textit{gymnasium's} frozen lake environment \cite{towers2024gymnasium} (right).}
\label{fig:maze}
\end{figure}

Gridworld games as depicted in  Fig. \ref{fig:maze} are particularly suitable for QRL benchmarking because their observation and action spaces naturally map to quantum encodings (one-hot or binary), and unlike Atari games, which have large state spaces that exceed current quantum capabilities, gridworld environments can be scaled appropriately. 

In the following we will therefore use established gridworlds from literature (\cite{sutton1990integrated}, \cite{Crawford.2018} and \cite{muller2021towards}) as well as \textit{gymnasium's} frozen lake environment for consistent benchmarking across different studies, addressing the current issue of fragmented, non-comparable results in QRL research. 

\subsection{Metrics}\label{metrics}

Evaluating and benchmarking QRL algorithms requires incorporating, but also expanding beyond traditional RL metrics. While current QRL works often focus solely on performance and sample efficiency, classical RL highlights the importance of overall clock time (e.g. in RL methods such as A3C through asynchronous parameter updates and multi-GPU usage  \cite{babaeizadeh2016reinforcement}). Therefore, we propose a set of five metrics, which will be analyzed across the benchmark environments: 
\begin{enumerate}
    \item \textit{performance}, assessing the algorithm's ability to achieve its objectives
    \item \textit{sample efficiency}, measuring the amount of environmental interaction required to reach a certain performance level (with a predefined maximal environment step limit)
    \item \textit{number of circuit executions}, highlighting the costliness of quantum computations
    \item \textit{quantum clock time}, influenced by circuit depth and quantum hardware
    \item \textit{qubit scaling}, crucial for estimating the future applicability of the approaches
\end{enumerate}

By examining QRL algorithms through these metrics, a better understanding of their strengths, weaknesses, and areas for improvement can be gained.

\subsection{Evaluating the Q in QRL}\label{eval-q}

The recent work of \cite{bowles2024better} has emphasized a question which has barely been investigated in QRL: Whether the observed advantages in quantum algorithms stem from quantum principles or other factors. In this Section we discuss how to answer this question for the analyzed classes of QRL algorithms.

\textbf{PQC-QRL:} PQCs have been the subject to detailed analyses. Current results suggest that the choice of ansatz is crucial in order to determine whether the PQC will suffer from untrainability (also called barren plateaus \cite{larocca2024review}) or be classically simulatable (as has recently been shown for quantum convolutional neural networks \cite{bermejo2024quantum}). To evaluate if the performance of the agents is due to quantum properties such as entanglement, we compare the original ansatz of \cite{skolik2022quantum} against two modified versions: For the first modified ansatz (A), we remove the entangling block, making the ansatz linearly separable, hence classically simulatable. For the second ansatz (B) we do the same but encode in each qubit the whole state space in the encoding block over the layers. By comparing the ansatz form \cite{skolik2022quantum} against these linearly separable ansatzes, we can access if the observed performance is due to the entanglement. 

\textbf{FE-QRL:} While QBM based FE-QRL has shown promise, with empirical results suggesting its potential to outperform classical DBM \cite{Crawford.2018} (also on D-Wave Quantum Annealers  \cite{levit2017free} and \cite{Neumann.2023}), several questions remain open. A significant challenge in FE-QRL is the approximation of the partition function, which is caused by the limitations of measuring spin configurations of qubits along a fixed axis. When a measurement of $\sigma_z$ is performed, the quantum state collapses into one of its eigenstates along the z-axis, irreversibly destroying any information about the spin's projection along the transverse fields direction (represented by $\sigma_x$). Therefore, it remains questionable, if a Quantum Annealer can be used to approximate Eq. \ref{eq:qbm} \cite{amin2015searching} \cite{matsuda2009ground} \cite{venuti2017relaxation}. As a result, previous works have introduced alternative methods to estimate $\langle H_v \rangle$. One widely adopted method is called \textit{replica stacking} \cite{levit2017free} \cite{Crawford.2018}, which utilizes the Suzuki-Trotter decomposition \cite{suzuki1976relationship} to construct an approximate Hamiltonian $H^{QBM'}_v$. Using the decomposition, the traverse field term of Eq. \ref{eq:qbm} is transformed into a classical Ising model of one dimension higher: 

\begin{equation}
\begin{split}
    H^{QBM'}_v = & - \sum_{\{h,h'\}\subseteq H} \sum^r_{k=1} \frac{\theta^{hh'}}{r}\sigma^z_{hk}\sigma^z_{h'k} \\ &- \sum_{v \in V, h \in H} \sum^r_{k=1} \frac{\theta^{vh}v}{r}\sigma^z_{hk} \\ &- w^+ \Big( \sum_{h \in H} \sum^r_{k=0} \sigma^z_{hk} \sigma^z_{hk+1} \Big),
\end{split}
\end{equation}

where $r$ is the number of \textit{replicas}, and $w^+ = \frac{1}{2\beta} log \ coth(\frac{\Gamma \beta }{r})$ \cite{levit2017free}.
\cite{suzuki1976relationship} shows, that as the amount of \textit{replicas} is increased, the ground state of $H^{QBM'}_v$ converges towards $H^{QBM}_v$. However, this does not imply $\langle H^{QBM}_v \rangle \approx  \langle H^{QBM'}_v \rangle $. Nevertheless, $H^{QBM'}_v$ is used throughout literature to approximate the free energy of the QBM. This is either done with Simulated Annealing, or via a D-Wave Quantum Annealer. Another problem arises from the unknown values of $\Gamma$ and $\beta$ in the approximation for $H^{QBM'}_v$. \cite{levit2017free} associate a single (average) virtual $\Gamma$ to all TFIMs constructed throughout the FE-QRL. While a validation of this approach is beyond the scope of this work, we will proceed with an empirical evaluation. This evaluation centers on the following hypothesis, drawn from the aforementioned studies: If $H^{QBM'}_v$ provides a good approximation of $H^{QBM}$, we expect superior training performance relative to the classical $H^{DBM}_v$. To test this hypothesis, we will examine if the performance of FE-QRL improves with an increasing number of \textit{replicas}. 

\textbf{AA-QRL:} For this QRL approach we do not conduct an additional analysis of its quantum principles, since in its evaluated form its referred to as QiRL (as discussed in Section \ref{sec:aa}).

\begin{figure}
    \centering
    \includegraphics[width=0.6\linewidth]{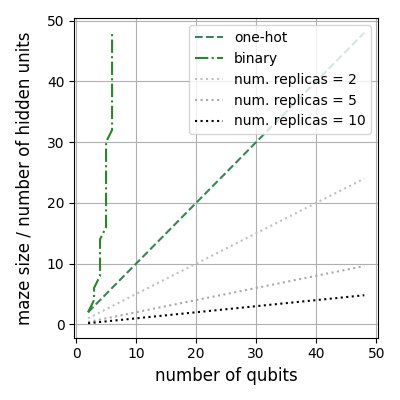}
    \caption{Number of required qubits: For PQC-QRL, the number of qubits greatly differs between binary and one-hot state space encoding. For FE-QRL, the number of qubits depends on the amount of hidden units of the QBM as well as the number of \textit{replicas} used for the approximation. }
    \label{fig:qubit_scaling}
\end{figure}

\begin{figure}
    \centering
    \includegraphics[width=1.0\linewidth]{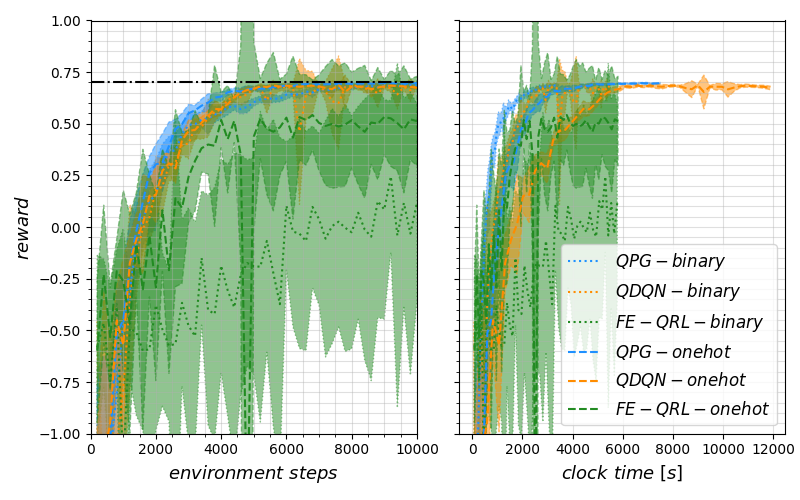}
    \caption{Comparison of different state space encodings on the $3\times3$ gridworld with an optimal reward of $0.7$, indicated by the dotted black line. The solid lines indicate the mean over 10 runs and the shaded area indicates the standard deviation.}
    \label{fig:binary_vs_onehot}
\end{figure}

\section{Results}

\begin{figure*}[h]
\centering
\begin{subfigure}{0.49\textwidth}
    \includegraphics[width=\textwidth]{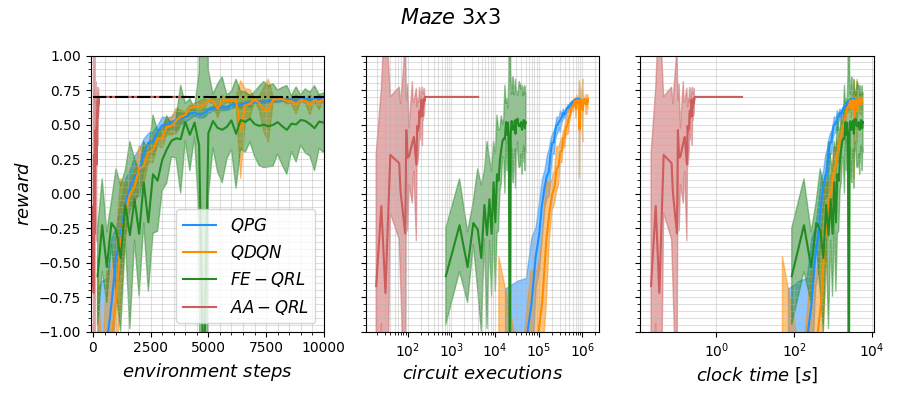}
    \label{fig:first}
\end{subfigure}
\hfill
\hfill
\begin{subfigure}{0.49\textwidth}
    \includegraphics[width=\textwidth]{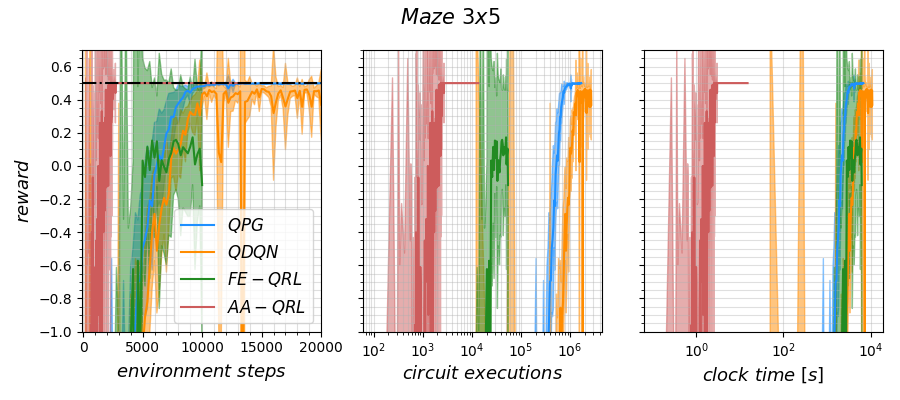}
    \label{fig:second}
\end{subfigure}
\hfill
\begin{subfigure}{0.49\textwidth}
    \includegraphics[width=\textwidth]{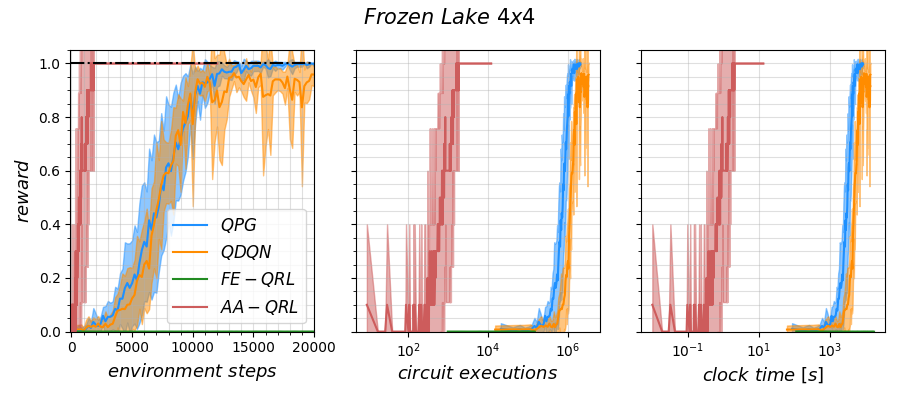}
    \label{fig:third}
\end{subfigure}
\hfill
\begin{subfigure}{0.49\textwidth}
    \includegraphics[width=\textwidth]{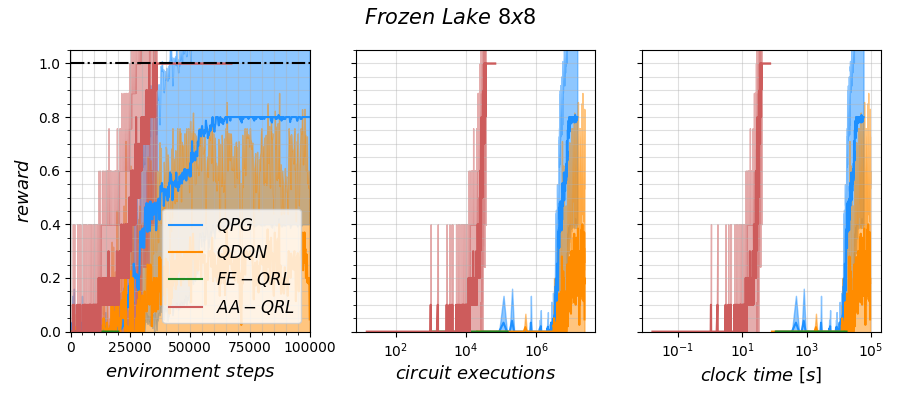}
    \label{fig:third}
\end{subfigure}
        
\caption{Comparison of the QRL algorithms on four gridworlds. Optimal rewards are indicated by the dotted black line. The solid lines show the mean over 10 runs and the shaded area the standard deviation.}
\label{fig:main}
\end{figure*}

We investigate the QRL agents proposed in Section \ref{algorithms} on the gridworld environments proposed in Section \ref{sec:benchmarks}: A simple $3\times3$ gridworld as proposed by \cite{muller2021towards}, a $3\times5$ gridworld as proposed by \cite{Crawford.2018} and an $4\times4$ as well as an $8\times8$ instance of the non-slippery frozen lake environment. The action space for all environments is chosen to be discrete with four possible actions (up,down,left,right), which are one-hot encoded for PQC-QRL and FE-QRL, and binary encoded for AA-QRL. 
For the PQC-QRL, we use PQCs with $4$, $5$, $7$ and $9$ qubits (depending on the state space of the environments) with $5$ layers each (as depicted in Fig. \ref{fig:vqc_hea}). For the FE-QRL approach we use two hidden layers with $4$ qubits each. To encode the four actions in the AA-QRL agent, two qubits are required. 

For QPG, we use a learning rate of $0.025$ for the parameters $\theta,\lambda$ and $0.1$ for the output scaling parameters $w$ for all environments. For the QDQN, we use a learning rate of $0.01$ for the parameters $\theta,\lambda$ and $0.01$ for the output scaling parameters $w$ for all environments, a $\gamma$ of $0.95$ and an epsilon decay rate from $1$ to $0.05$. For the simulation of the PQCs we use state vector simulators.

For the FE-QRL agents, the choice of hyperparameters is extremely important. Throughout our experiments, slight modification lead to strong fluctuations in performance. Since we do not want to bias our evaluation and fine tune the algorithms significantly more than the other algorithms, we do use learning rate schedules (as proposed by \cite{Crawford.2018}), but do not fine tune them for the different gridworld environments. Additionally, we use $\beta=2.0$, $\Gamma=0.506$, and the same $\gamma$ and epsilon greedy exploration schedule as for the QDQN agents. To estimate $\langle H^{QBM'}_v \rangle$ we use Simulated Annealing.

As discussed in Section \ref{metrics}, we evaluate not only the performance of the agents in terms of environment interactions but also with respect to the amount of circuit executions and quantum clock time. The number of circuit executions is influenced by both the number of forward passes and the number of model parameters, particularly in the case of PQC-QRL, since the parameter-shift rule necessitates a minimum of two circuit executions per parameter for gradient estimation. Consequently, QDQN agents require a higher number of circuit executions for the same amount of environment interactions compared to QPG, as Q-Learning  (in our implementation) revisits previously seen data through resampling from the replay buffer more often. For PQC-QRL and AA-QRL, the estimated quantum clock time is derived from assumed gate times on superconducting hardware of $30ns$ and $300ns$ for single and two-qubit gates, respectively, as well as $300ns$ measurement times, with $1000$ shots. The estimated quantum clock time for FE-QRL is based on the usage of the D-Wave Quantum Annealer Advantage QPU. A single $4\times4$ QBM, approximated with $5$ \textit{replicas} and with the default anneal schedule and $1000$ shots requires approximately $115ms$ of QPU access time.

An important question is whether QRL agents can scale to larger problem instances. To answer this question, we need to consider the qubit scaling for the different approaches. When using one-hot encoding for the state space, the PQC-QRL faces significant scalability issues, since the number of qubits scales linearly with the size of the state space (ref. Fig. \ref{fig:qubit_scaling}). For binary encoding on the other hand, this scaling is significantly better.  Note that one needs at least $4$ qubits for the one hot encoding of the actions. In contrast,  FE-QRL's number of qubits is unaffected by the encoding method, since the state is represented via the visible nodes. However, a higher number of visible nodes (due to the use of one-hot encoding) leads to a higher number of trainable parameters. The AA-QRL is insensitive to the encoding scheme, as a separate quantum circuit is employed for each state, making the number of required qubits independent of the encoding. 

In Fig. \ref{fig:binary_vs_onehot} the comparison of the PQC-QRL algorithms with binary encoding ($4$ qubits, $5$ layers) and with one-hot encoding ($9$ qubits, $5$ layers) shows that even though the number of trainable parameters is more than twice as high, the performance is comparable in terms of environmental steps. However, due to the increased number of parameters, the performance of the larger models is worse in terms of quantum clock time. On the other hand, the binary encoding for the FE-QRL agent performs significantly worse than the one-hot encoded agent, while the clock time remains the same, since the different encodings only affect the visible nodes. 

In the comparison in Fig. \ref{fig:main} we therefore evaluate the agents with binary encoding for the PQC-agents and the one-hot encoded for the FE-QRL agents. Throughout all gridworlds, the AA-QRL method performs best. This becomes especially apparent for the quantum clock time of the algorithm. However, as the size of the gridworlds grow, the method seems to require proportionally more environment steps (compared to QPG and QDQN). The performance of the QPG agent and the QDQN agent is similar throughout the small gridworld sizes. However, for the largest frozen lake gridworld, the performance is QDQN starts to deteriorate. The FE-QRL agents are incapable of scaling to the larger frozen lake environments. While the method has shown promising results in \cite{Crawford.2018} and other works, it does not seem to scale well to larger problem instances. The number of circuit executions is less for the FE-QRL agents than for the QDQN and QPG agent, but due to the longer quantum clock times of a single circuit executions, the overall quantum clock times of the two approaches is comparable. 

While all algorithms have the potential to scale up to problem sizes far beyond the ones utilized in this work, their performance greatly decreases as problem sizes grow. 

\begin{figure}
\centering
 \includegraphics[width=0.8\linewidth]{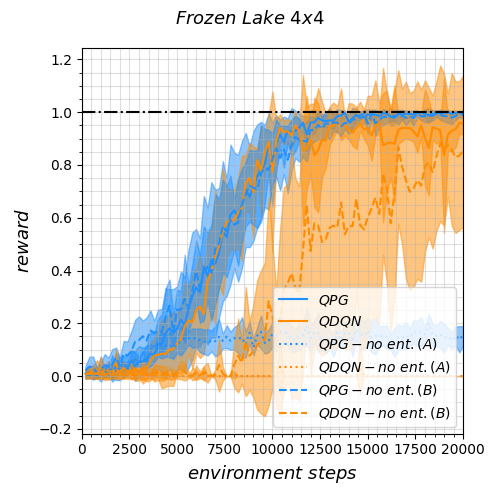}
\caption{Performance of PQC-QRL algorithms with and without entanglement on the frozen lake $4\times4$. }\label{fig:no_ent}
\end{figure}

In order to access if the performance of PQC-QRL relies on quantum properties such as entanglement, we compare the performance of the proposed ansatz by \cite{skolik2022quantum} against ansatze without any entangling gates, effectively removing the entangling blocks (ref. Fig. \ref{fig:vqc_hea}). The results in Fig. \ref{fig:no_ent} show that the first models without entanglement (A) perform significantly worse. This is due to a lack of information encoding on the individual qubits. The second model without entanglement (B) includes all information on each qubit, and performs almost identical for the QDQN agents, but worse for the QPG agents. The linearly separable ansatze show only partly worse training performance, and hence the performance of the quantum algorithm does not seem to mainly rely on entanglement.

\begin{figure}
\centering

\includegraphics[width=0.8\linewidth]{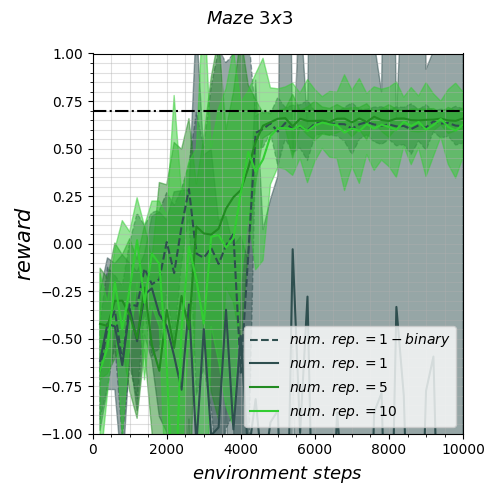}
\caption{Performance of FE-QRL with increasing number of \textit{replicas} and binary and onehot encoding on the $3\times3$ gridworld.}\label{fig:replicas}
\end{figure}

We compare the performance of the FE-QRL approach for an increasing numbers of \textit{replicas} via simulated Annealing: 1 (so a classical DBM), 5 and 10. We also evaluate the classical DBM with binary and onehot encoding.  As discussed in Section \ref{eval-q}, we would expect that an increase of the number of \textit{replicas} would result in better training performance, if the performance of the QBM relies on quantum principles. However, as we can see in Fig. \ref{fig:replicas}, we see no such correlation. While the set of hyperparameters does not lead to good performance for the onehot encoded classical DBM, the performance of the FE-QRL with 5 and 10 replicas is comparable to the binary encoded DBM. Hence, the performance of the FE-QRL approach seems to rely too strongly on hyperparameters, making a meaningful ranking unfeasible.

\section{Discussion}

In this study, we conducted a comprehensive evaluation of three QRL classes (PQC-QRL with QPG and QDQN, FE-QRL and AA-QRL). Our evaluation extends beyond previous works by the number of considered QRL algorithms and the incorporation of additional metrics such as circuit executions and quantum clock time, providing a more holistic and realistic assessment of these algorithms' practical feasibility.

For PQC-QRL, we observed only a minor dependence on quantum entanglement, with performance deteriorating only slightly when entanglement was removed. Interestingly, our investigation of FE-QRL showed no clear correlation between performance and the number of \textit{replicas} used to approximate the Hamiltonian of the QBM $H^{QBM}_v$, but rather a great dependence on hyperparameters. These findings suggest that most QRL approaches may not greatly rely on their quantum components. 

QRL, particularly when applied to gridworld games, demonstrates promising scalability to larger problems through binary encoding, even with current hardware limitations. However, the algorithms we evaluated still require substantial improvement to achieve competitive performance levels. Our work can serve as an underlying benchmarking reference for this future development. 

Future work should aim to include the evaluation of noise resilience as an additional metric in order to assess these algorithms' practical viability in real quantum hardware implementations. Additionally, not only the quantum clock time, but also the overall clock time of these hybrid algorithms should be considered when comparing QRL to classical RL.

\subsubsection*{\uppercase{Code Availability}}
The code to reproduce the results as well as the data used to generate the plots in this work can be found here: https://github.com/georgkruse/cleanqrl

\subsubsection*{\uppercase{Statement of Independent Work}}
We acknowledge the existence of a research paper with the same title as ours \cite{meyer2025benchmarking}. We wish to clarify that both works were conducted independently and without knowledge of each other.

\subsubsection*{\uppercase{Acknowledgements}}

The research is part of the Munich Quantum Valley, which
is supported by the Bavarian state government with funds from
the Hightech Agenda Bayern Plus.

\bibliographystyle{apalike}
{\small
\bibliography{example}}

\begin{thebibliography}{}

\bibitem[Abbas et~al., 2021]{abbas2021power}
Abbas, A., Sutter, D., Zoufal, C., Lucchi, A., Figalli, A., and Woerner, S.
  (2021).
\newblock The power of quantum neural networks.
\newblock {\em Nature Computational Science}, 1(6):403--409.

\bibitem[Ackley et~al., 1985]{ackley1985learning}
Ackley, D.~H., Hinton, G.~E., and Sejnowski, T.~J. (1985).
\newblock A learning algorithm for boltzmann machines.
\newblock {\em Cognitive science}, 9(1):147--169.

\bibitem[Amin, 2015]{amin2015searching}
Amin, M.~H. (2015).
\newblock Searching for quantum speedup in quasistatic quantum annealers.
\newblock {\em Physical Review A}, 92(5):052323.

\bibitem[Amin et~al., 2018]{Amin.2018}
Amin, M.~H., Andriyash, E., Rolfe, J., Kulchytskyy, B., and Melko, R. (2018).
\newblock Quantum boltzmann machine.
\newblock {\em Physical Review X}, 8(2):0541.

\bibitem[Babaeizadeh et~al., 2016]{babaeizadeh2016reinforcement}
Babaeizadeh, M., Frosio, I., Tyree, S., Clemons, J., and Kautz, J. (2016).
\newblock Reinforcement learning through asynchronous advantage actor-critic on
  a gpu.
\newblock {\em arXiv preprint arXiv:1611.06256}.

\bibitem[Bermejo et~al., 2024]{bermejo2024quantum}
Bermejo, P., Braccia, P., Rudolph, M.~S., Holmes, Z., Cincio, L., and Cerezo,
  M. (2024).
\newblock Quantum convolutional neural networks are (effectively) classically
  simulable.
\newblock {\em arXiv preprint arXiv:2408.12739}.

\bibitem[Bowles et~al., 2024]{bowles2024better}
Bowles, J., Ahmed, S., and Schuld, M. (2024).
\newblock Better than classical? the subtle art of benchmarking quantum machine
  learning models.
\newblock {\em arXiv preprint arXiv:2403.07059}.

\bibitem[Chen et~al., 2020]{chen2020variational}
Chen, S. Y.-C., Yang, C.-H.~H., Qi, J., Chen, P.-Y., Ma, X., and Goan, H.-S.
  (2020).
\newblock Variational quantum circuits for deep reinforcement learning.
\newblock {\em IEEE access}, 8:141007--141024.

\bibitem[Coelho et~al., 2024]{coelho2024vqc}
Coelho, R., Sequeira, A., and Paulo~Santos, L. (2024).
\newblock Vqc-based reinforcement learning with data re-uploading: performance
  and trainability.
\newblock {\em Quantum Machine Intelligence}, 6(2):53.

\bibitem[Crawford et~al., 2018]{Crawford.2018}
Crawford, D., Levit, A., Ghadermarzy, N., Oberoi, J.~S., and Ronagh, P. (2018).
\newblock Reinforcement learning using quantum boltzmann machines.
\newblock {\em Quantum Information {\&} Computation}.

\bibitem[Dong et~al., 2010]{dong2010robust}
Dong, D., Chen, C., Chu, J., and Tarn, T.-J. (2010).
\newblock Robust quantum-inspired reinforcement learning for robot navigation.
\newblock {\em IEEE/ASME transactions on mechatronics}, 17(1):86--97.

\bibitem[Dong et~al., 2008]{dong2008quantum}
Dong, D., Chen, C., Li, H., and Tarn, T.-J. (2008).
\newblock Quantum reinforcement learning.
\newblock {\em IEEE Transactions on Systems, Man, and Cybernetics, Part B
  (Cybernetics)}, 38(5):1207--1220.

\bibitem[Dr{\u{a}}gan et~al., 2022]{druagan2022quantum}
Dr{\u{a}}gan, T.-A., Monnet, M., Mendl, C.~B., and Lorenz, J.~M. (2022).
\newblock Quantum reinforcement learning for solving a stochastic frozen lake
  environment and the impact of quantum architecture choices.
\newblock {\em arXiv preprint arXiv:2212.07932}.

\bibitem[Hu et~al., 2021]{hu2021quantum}
Hu, Y., Tang, F., Chen, J., and Wang, W. (2021).
\newblock Quantum-enhanced reinforcement learning for control: A preliminary
  study.
\newblock {\em Control Theory and Technology}, 19:455--464.

\bibitem[Jerbi et~al., 2021a]{jerbi2021parametrized}
Jerbi, S., Gyurik, C., Marshall, S., Briegel, H., and Dunjko, V. (2021a).
\newblock Parametrized quantum policies for reinforcement learning.
\newblock {\em Advances in Neural Information Processing Systems},
  34:28362--28375.

\bibitem[Jerbi et~al., 2021b]{Jerbi.2021}
Jerbi, S., Trenkwalder, L.~M., {Poulsen Nautrup}, H., Briegel, H.~J., and
  Dunjko, V. (2021b).
\newblock Quantum enhancements for deep reinforcement learning in large spaces.
\newblock {\em PRX Quantum}, 2(1).

\bibitem[Kappen, 2020]{kappen2020learning}
Kappen, H.~J. (2020).
\newblock Learning quantum models from quantum or classical data.
\newblock {\em Journal of Physics A: Mathematical and Theoretical},
  53(21):214001.

\bibitem[Kruse et~al., 2024]{kruse2024hamiltonian}
Kruse, G., Coehlo, R., Rosskopf, A., Wille, R., and Lorenz, J.~M. (2024).
\newblock Hamiltonian-based quantum reinforcement learning for neural
  combinatorial optimization.
\newblock {\em arXiv preprint arXiv:2405.07790}.

\bibitem[Kruse et~al., 2023]{kruse2023variational}
Kruse, G., Dragan, T.-A., Wille, R., and Lorenz, J.~M. (2023).
\newblock Variational quantum circuit design for quantum reinforcement learning
  on continuous environments.
\newblock {\em arXiv preprint arXiv:2312.13798}.

\bibitem[Larocca et~al., 2024]{larocca2024review}
Larocca, M., Thanasilp, S., Wang, S., Sharma, K., Biamonte, J., Coles, P.~J.,
  Cincio, L., McClean, J.~R., Holmes, Z., and Cerezo, M. (2024).
\newblock A review of barren plateaus in variational quantum computing.
\newblock {\em arXiv preprint arXiv:2405.00781}.

\bibitem[Levit et~al., 2017]{levit2017free}
Levit, A., Crawford, D., Ghadermarzy, N., Oberoi, J.~S., Zahedinejad, E., and
  Ronagh, P. (2017).
\newblock Free energy-based reinforcement learning using a quantum processor.
\newblock {\em arXiv preprint arXiv:1706.00074}.

\bibitem[Matsuda et~al., 2009]{matsuda2009ground}
Matsuda, Y., Nishimori, H., and Katzgraber, H.~G. (2009).
\newblock Ground-state statistics from annealing algorithms: quantum versus
  classical approaches.
\newblock {\em New Journal of Physics}, 11(7):073021.

\bibitem[Meyer et~al., 2022]{meyer2022survey}
Meyer, N., Ufrecht, C., Periyasamy, M., Scherer, D.~D., Plinge, A., and
  Mutschler, C. (2022).
\newblock A survey on quantum reinforcement learning.
\newblock {\em arXiv preprint arXiv:2211.03464}.

\bibitem[Meyer et~al., 2025]{meyer2025benchmarking}
Meyer, N., Ufrecht, C., Yammine, G., Kontes, G., Mutschler, C., and Scherer,
  D.~D. (2025).
\newblock Benchmarking quantum reinforcement learning.
\newblock {\em arXiv preprint arXiv:2501.15893}.

\bibitem[M{\"u}ller et~al., 2021]{muller2021towards}
M{\"u}ller, T., Roch, C., Schmid, K., and Altmann, P. (2021).
\newblock Towards multi-agent reinforcement learning using quantum boltzmann
  machines.
\newblock {\em arXiv preprint arXiv:2109.10900}.

\bibitem[Neumann et~al., 2023]{Neumann.2023}
Neumann, N. M.~P., de~Heer, P. B. U.~L., and Phillipson, F. (2023).
\newblock Quantum reinforcement learning - comparing quantum annealing and
  gate-based quantum computing with classical deep reinforcement learning.
\newblock {\em Quantum Information Processing}, 22(2).

\bibitem[Sallans and Hinton, 2004]{sallans2004reinforcement}
Sallans, B. and Hinton, G.~E. (2004).
\newblock Reinforcement learning with factored states and actions.
\newblock {\em The Journal of Machine Learning Research}, 5:1063--1088.

\bibitem[Skolik et~al., 2022]{skolik2022quantum}
Skolik, A., Jerbi, S., and Dunjko, V. (2022).
\newblock Quantum agents in the gym: a variational quantum algorithm for deep
  q-learning.
\newblock {\em Quantum}, 6:720.

\bibitem[Skolik et~al., 2023]{skolik2023robustness}
Skolik, A., Mangini, S., B{\"a}ck, T., Macchiavello, C., and Dunjko, V. (2023).
\newblock Robustness of quantum reinforcement learning under hardware errors.
\newblock {\em EPJ Quantum Technology}, 10(1):1--43.

\bibitem[Sutton, 1990]{sutton1990integrated}
Sutton, R.~S. (1990).
\newblock Integrated architectures for learning, planning, and reacting based
  on approximating dynamic programming.
\newblock In {\em Machine learning proceedings 1990}, pages 216--224. Elsevier.

\bibitem[Suzuki, 1976]{suzuki1976relationship}
Suzuki, M. (1976).
\newblock Relationship between d-dimensional quantal spin systems and (d+
  1)-dimensional ising systems: Equivalence, critical exponents and systematic
  approximants of the partition function and spin correlations.
\newblock {\em Progress of theoretical physics}, 56(5):1454--1469.

\bibitem[Towers et~al., 2024]{towers2024gymnasium}
Towers, M., Kwiatkowski, A., Terry, J., Balis, J.~U., De~Cola, G., Deleu, T.,
  Goulao, M., Kallinteris, A., Krimmel, M., KG, A., et~al. (2024).
\newblock Gymnasium: A standard interface for reinforcement learning
  environments.
\newblock {\em arXiv preprint arXiv:2407.17032}.

\bibitem[Venuti et~al., 2017]{venuti2017relaxation}
Venuti, L.~C., Albash, T., Marvian, M., Lidar, D., and Zanardi, P. (2017).
\newblock Relaxation versus adiabatic quantum steady-state preparation.
\newblock {\em Physical Review A}, 95(4):042302.

\end{thebibliography}

\end{document}